\documentclass{aastex631}

\usepackage{graphicx,subfigure}
\usepackage[utf8]{inputenc}
\usepackage{longtable}
\usepackage{booktabs}
\usepackage{multirow}
\usepackage{savesym}
\savesymbol{tablenum}
\usepackage{siunitx}
\restoresymbol{SIX}{tablenum}
\usepackage{natbib}
\usepackage{amsmath}
\graphicspath{{./}{figures/}}
\DeclareUnicodeCharacter{2009}{\,}

\begin{document}

\title{New resonance scattering model in AtomDB: application to line suppression in galaxy clusters and elliptical galaxies}

\author[0000-0002-4469-2518]{Priyanka Chakraborty}
\affiliation{Center for Astrophysics $\vert$ Harvard \& Smithsonian \\
}
\email{priyanka.chakraborty@cfa.harvard.edu}
\author[0000-0003-3462-8886]{Adam Foster}
\affiliation{Center for Astrophysics $\vert$ Harvard \& Smithsonian \\
}
\author[0000-0003-4284-4167]{Randall Smith}
\affiliation{Center for Astrophysics $\vert$ Harvard \& Smithsonian \\
}
\author[0000-0002-8704-4473]{Nancy Brickhouse}
\affiliation{Center for Astrophysics $\vert$ Harvard \& Smithsonian \\
}
\author[0000-0002-7868-1622]{John Raymond}
\affiliation{Center for Astrophysics $\vert$ Harvard \& Smithsonian \\
}

\begin{abstract}
In this paper, we present a simple, one-step, self-consistent, and fast resonance scattering model \texttt{rsapec} based on the AtomDB database. This model can be used as an alternative to the commonly used APEC model for fitting such X-ray spectra with optically thick lines. The current model is intended, in general, for verifying the presence of the effect and for spectral modeling of galaxy clusters and elliptical galaxies under applicable assumptions. We test \texttt{rsapec} to derive the line suppression in the elliptical galaxy NGC 4636 and the Perseus cluster of galaxies, and obtain resonance suppression of $\sim$ 1.24 and $\sim$ 1.30, respectively.

\end{abstract}

\section{Introduction:}

X-ray spectral lines are subject to being  distorted by resonant scattering (RS). Resonant scattering occurs when optically thick line photons are absorbed by ions with similar resonant transition energies and are
immediately re-emitted  in another direction. This process redirects photons, suppressing resonance line emission in some areas with matching enhancements in others. For example, in galaxies and clusters of galaxies, the resonant line intensities 
are reduced towards their center and enhanced towards their outskirts.

Over the years, several studies reported the detection of the RS effect and demonstrated the importance of incorporating resonance scattering in modeling the X-ray spectra \citep{1987SvAL...13....3G, 1998ApJ...497..587S, 2004MNRAS.347...29C, 2006MNRAS.370...63S, 2023MNRAS.522.3665N}.  RS is crucial for the accurate determination of the abundances of heavy elements  in the intra-cluster medium \citep{2000AdSpR..25..603A, 2022MNRAS.516.3068S},  measurements of velocity fields and anisotropy of gas motion \citep{2010SSRv..157..193C, 2022ApJ...935L..23S, 2023ApJ...944..132S}, and the polarization of bright X-ray emission lines in galaxy clusters and elliptical galaxies \citep{2010MNRAS.403..129Z}. RS can be important
for diagnostics of density or ionization state using He-like ions, Fe L shell ions, or the ratios of
He-like to H-like ions \citep{1992ApJ...398L.115S}. RS
has  also been suggested as an explanation for the weakness of O VII lines in the
shell of the Cygnus Loop supernova remnant, which otherwise would
imply a low oxygen abundance or charge transfer \citep{2019ApJ...871..234U}.

Tackling the problem of RS is not straightforward. Monte Carlo radiative transfer simulations have been extensively  used in the context of RS in galaxies and clusters of galaxies \citep{1987SvAL...13....3G, 2002MNRAS.333..191S, 2006ApJ...643L..73M, 2011AstL...37..141Z}. Examples include  the \texttt{Geant4} toolkit used for simulating the passage of photons through matter with functionalities like complex geometries and tracking \citep{2003NIMPA.506..250A, 2018PASJ...70...10H} and radiative Monte Carlo simulations used for the cluster ICM \citep{2002MNRAS.333..191S, 2004MNRAS.347...29C, 2010MNRAS.403..129Z}. 
The above techniques are rigorous and time-consuming.
 
In this paper, we present a simple, one-step,  and fast RS model for collisionally-ionized plasma for elliptical galaxies and galaxy clusters. Our new RS model, \texttt{rsapec}, can be used as an alternative to the commonly used APEC model for fitting the X-ray spectrum with optically thick lines. 
This model aims to verify the presence of the effect and for spectral modeling of galaxy clusters and elliptical galaxies under relevant assumptions outlined in \citet{1998ApJ...497..587S}.
Sections \ref{s:2} and \ref{s:3} elaborate on the theoretical background and the model parameters used for constructing the \texttt{rsapec} model. Section \ref{s:4} shows the application of the \texttt{rsapec} model to the elliptical galaxy NGC 4636 and the Perseus cluster of galaxies. Section 5 discusses our results.


\section{Theoretical Background:}\label{s:2}

\subsection{What drives intensity change in resonance lines?}

Line intensities in observed spectra is suppressed if line photons are scattered or destroyed. 
In resonance scattering, photons in the lines considered are not destroyed but just scattered out of the line of sight, effectively causing a migration of photons from the dense to the more diffuse regions of astrophysical objects.  This model does not include photons scattered into the line of sight, which
is only important in the outer parts of a cluster and can even enhance the
resonance line strength \citep{1987SvAL...13....3G}. Resonance enhancement will be addressed in a future paper.

Photons can also be scattered due to electron scattering escape, as a result of photons scattering off high-speed electrons, and receiving large Doppler-shifts from their line-center \citep{2021ApJ...912...26C, 2022ApJ...935...70C}. In addition, photons can be destroyed due to processes like Case A (optically thin) to Case B (optically thick) transfer and resonant Auger destruction \citep{1996MNRAS.278.1082R, 2020RNAAS...4..184C, 2020ApJ...901...68C, 2020ApJ...901...69C}. While  the above processes become important at large hydrogen column densities (N$_{H}$ $\sim$ 10$^{23}-$10$^{24}$ cm$^{-2}$), RS effects can noticeably alter line intensities at the much smaller hydrogen column densities (N$_{H}$ $\sim$ 10$^{20}-$10$^{21}$ cm$^{-2}$) typically found in galaxy clusters and elliptical galaxies considered in this paper.



\subsection{Calculating the RS factor}
Optical depth at a line's center ($\tau_{0}$) is defined as:

\begin{equation}\label{e:gamma1}
{\tau_{0}} = \int_{{0}}^\infty \frac{{\sqrt{ \pi}} h r_{e} cf n_{p} Z \delta_{i}}{\Delta E}dr
\end{equation}

where $h$ is the Planck constant, $c$ is the speed of light, $r_{e}$ is the classical electron radius, f is the oscillator strength for the given transition, $n_{p}$ is the proton number density, $Z$ is the elemental abundance, $\delta_{i}$ is the ion fraction at the given temperature, $\Delta E$ is the Doppler velocity width described by the following equation:

\begin{equation}\label{e:gamma2}
{\Delta E} = E_{0} \left(\frac{2kT}{Am_{p}c^{2}} + 2\frac{V_{1,turb}}{c^{2}}^{2}\right)^{1/2}
\end{equation}

where $E_{0}$ is the energy of the transition, $A$ is the atomic weight of the element, $m_{p}$ is the proton mass, and  $V_{1,turb}$ is the the line of sight velocity dispersion \citep{2018SSRv..214..108G}.

We adopt a similar approach as used by \citet{1998ApJ...497..587S}  for calculating the emergence of resonance lines of line-center optical depth $\tau_{0}$ from  optically thick plasma. 
 \citet{1998ApJ...497..587S}  investigated the effect of resonance line scattering in altering the surface brightness profile of resonance line emissions by numerically solving  radiative transfer equations for an isotropically emitting radiation source by $\Lambda$-iteration method , and estimated the RS factor ($f$, spectral intensity ratio with and without resonance scattering) at the line-center (see figure 4 in their paper).

We use their reported $f$ values and interpolated for the optical depths which didn't have the information for the RS factor.
For computational efficiency, we exclude lines with $\tau$ $<$ 0.05 with negligible line scattering ($<$ 2\%) from the RS calculation. The fraction of photons scattered away due to RS at the line-center  is therefore f$_{scattered}$= 1-$f$.

Line optical depths vary along the line profile following a Voigt profile distribution. Line  photons are absorbed and re-emitted, diffusing slowly in space until they reach the wings of the line with small optical depth and freely escape the cloud \citep{1962ApJ...135..195O}. For simplicity, it can be assumed that the scattered photons follow the Doppler line profile $\alpha (x)$=
exp(-x$^{2}$), where x =(E - E$_{0}$)/{$\Delta$ E} \citep{1983ASPRv...2..189P}. This Gaussian approximation is sufficient given the resolving power and signal-to-noise of the current and near-future X-ray observatories.
We construct a Gaussian line profile G(x) =   I$_{0}$  f$_{scattered}$  exp(-x$^{2}$) to account for the photons scattered away from our line of sight, where I$_{0}$ is the maximum intensity of the unscattered default AtomDB line profile.
We further subtract G(x) from the default AtomDB line profiles to produce the RS-corrected line profiles.


\begin{figure}[h!]
\centering
\subfigure{\includegraphics[width = 3in]{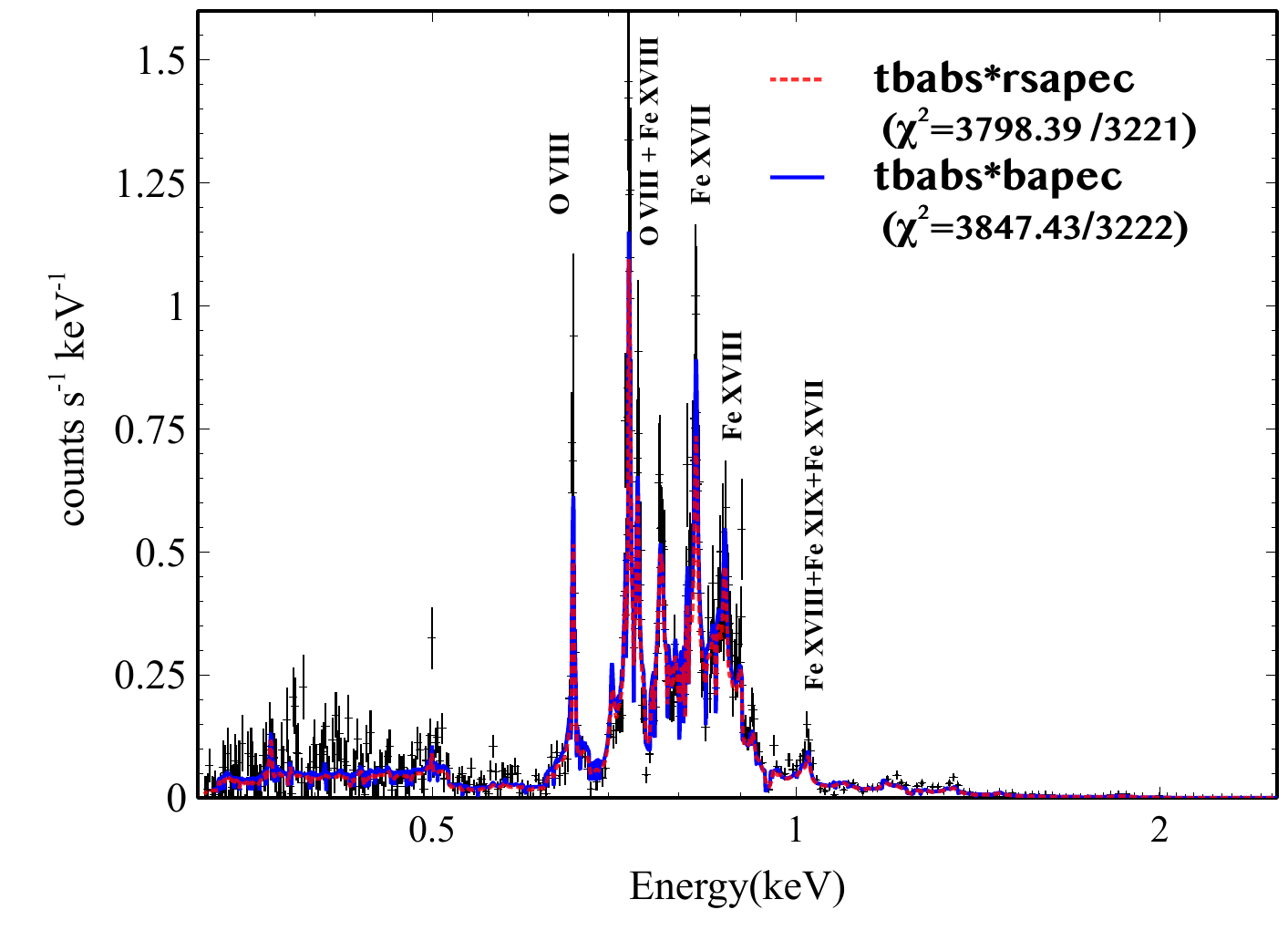}}\hspace{2cm}
\subfigure{\includegraphics[width = 3in]{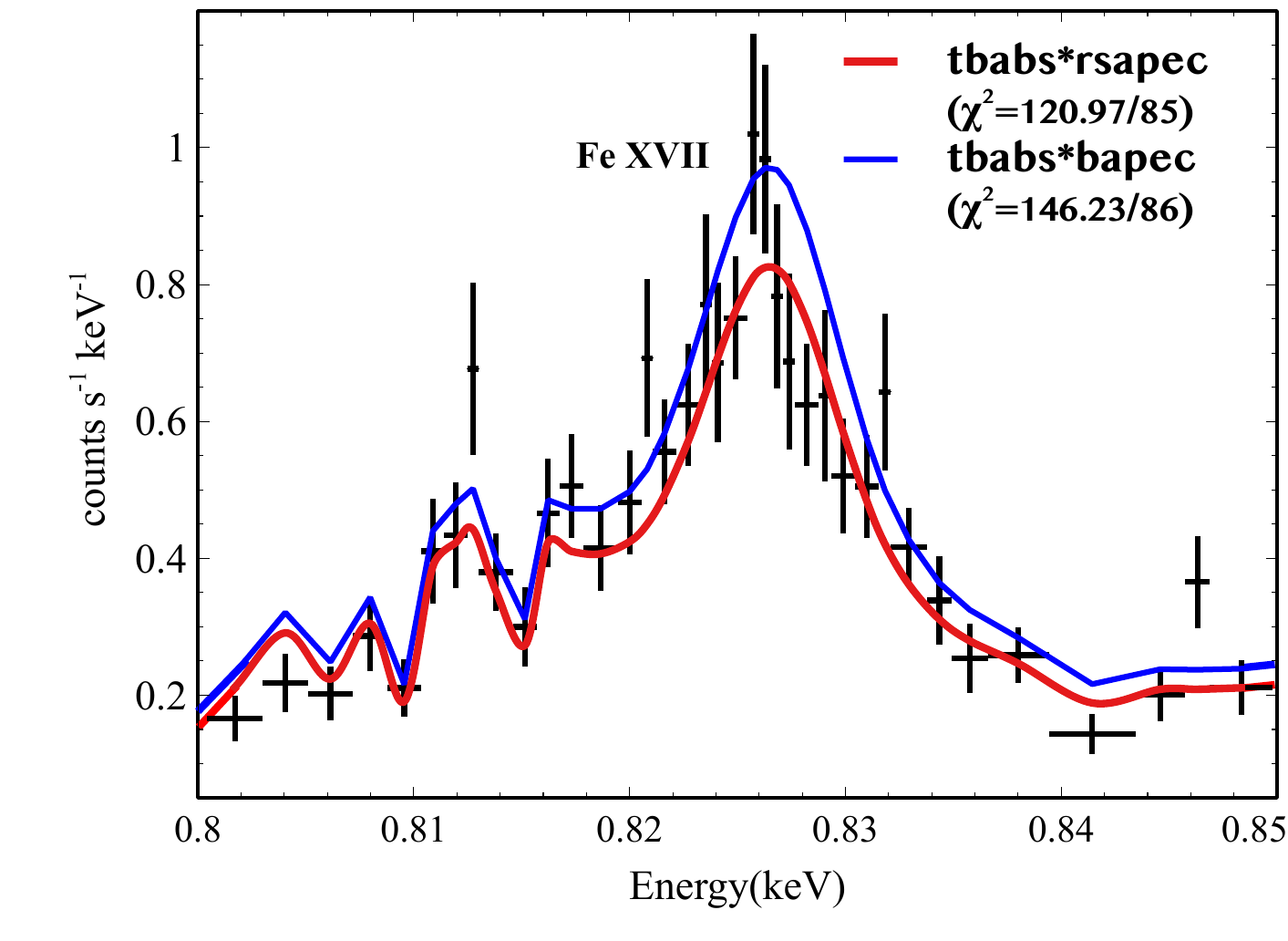}} 
\caption{Top: X-ray spectrum of the elliptical galaxy NGC 4636 observed by RGS/XMM-Newton within the energy range 0.33-2.5 keV using m$_{1}$(=\texttt{tbabs*bapec}, blue line) and m$_{2}$ (\texttt{=tbabs*rsapec}, red line) Bottom: The same spectrum fitted with m$_{1}$ and m$_{2}$ within a narrower energy range (0.8-0.85 keV) around the optically thick Fe XVII line.}
\label{f:2}
\end{figure}

\begin{figure}
\subfigure{\includegraphics[width = 3in]{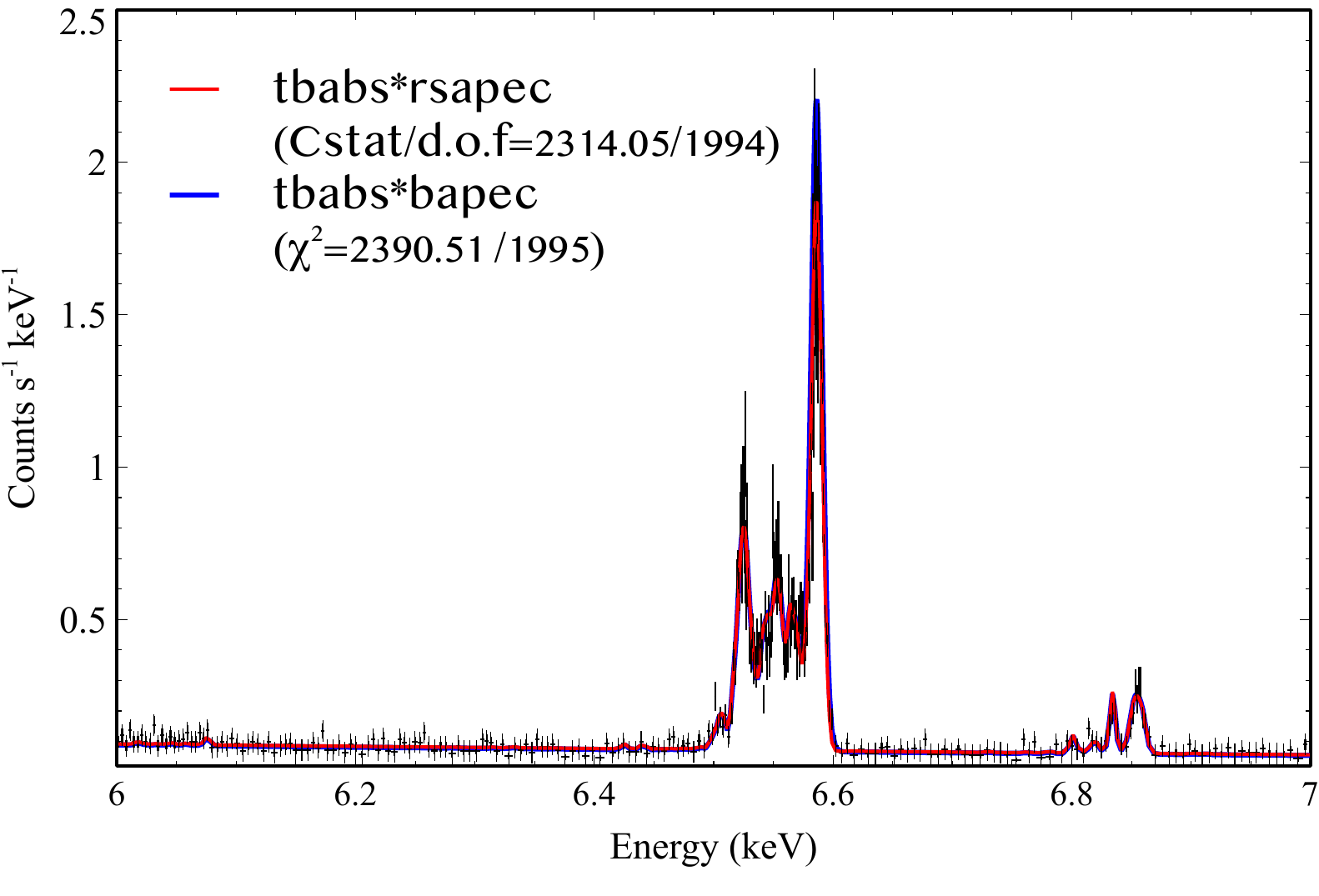}}\hspace{2cm}
\subfigure{\includegraphics[width = 3in]{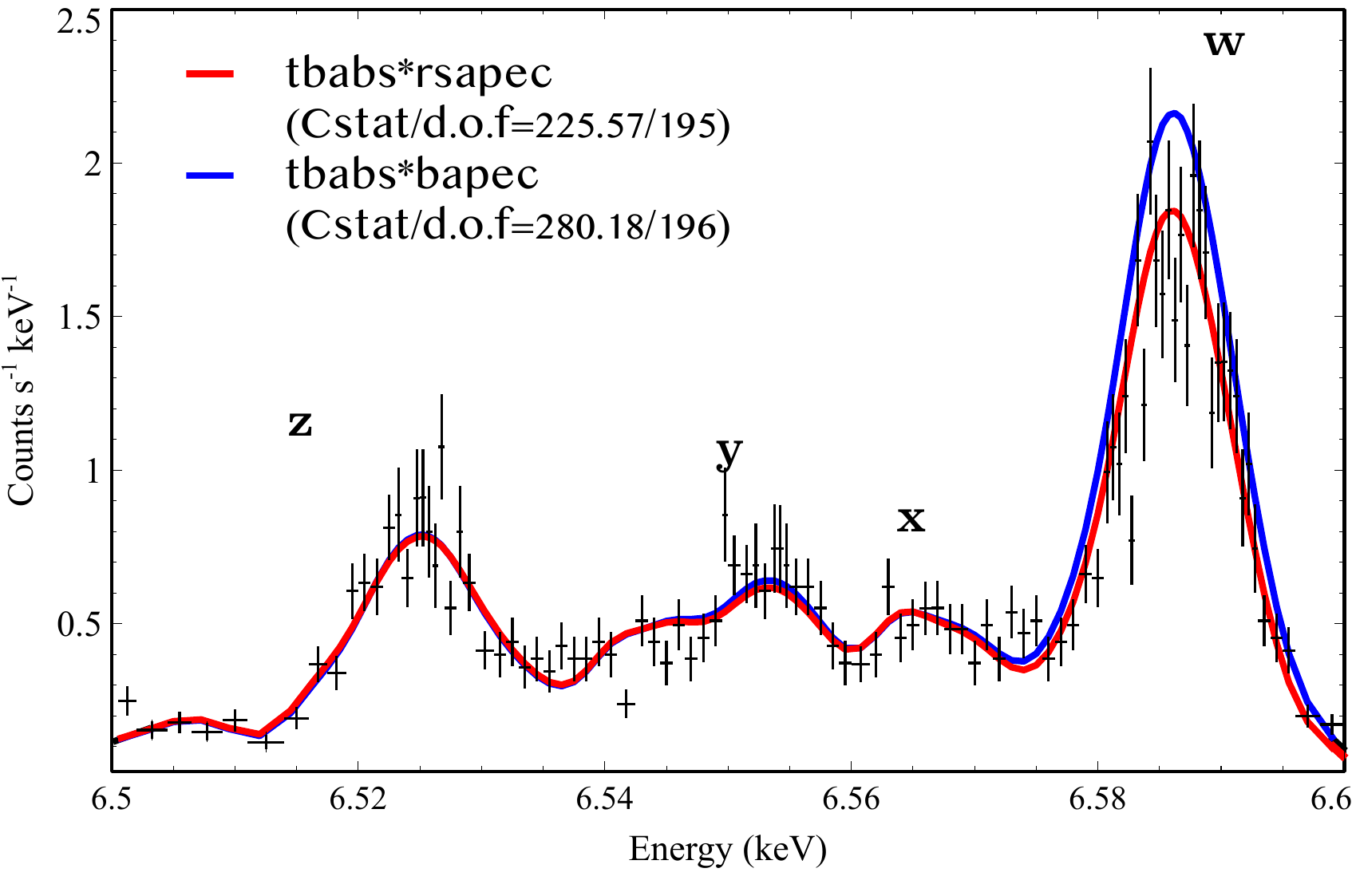}} 
\caption{Left: X-ray spectrum of the Perseus cluster core (outer region) 
 fitted with m$_{1}$(=\texttt{tbabs*bapec}, blue line) and m$_{2}$ (\texttt{=tbabs*rsapec}, red line) within the energy range 6.0-7.0 keV. Right: The same spectrum fitted with m$_{1}$ and m$_{2}$ within a narrower energy range (6.5-6.6 keV) around the Fe K$\alpha$ complex.}
\label{fig:FeXXIII_FeXXIV_FeXXV_FeXXVI}
\end{figure}

\begin{table*}
\centering{
\caption{\label{t:o1} Archival  RGS/\textit{XMM-Newton} observation of NGC 4636 and SXS/\textit{Hitomi} observation of Perseus cluster. The SXS data for 4 observations (observation  id: 100040020–100040050) were combined
for a total exposure time of 243.53 ksec.}}
\begin{tabular}{ccccc}
\hline
Instrument & Observation ID  & Start Date/Time  &  Exposure(ksec)\\
\hline
RGS/XMM-Newton & 0111190701 & 2000-07-13/03:44:33 & 64.41 \\
\hline
 & 100040020 & 2016-02-25/02:14:12  & 97.44 \\
 & 100040030 & 2016-03-04/02:17:32  & 72.51 \\
SXS/Hitomi & 100040040 & 2016-03-05 12:00:15  & 68.13 \\
 & 100040050 & 2016-03-06 19:37:59  & 5.45 \\
 \hline
\end{tabular}
\end{table*}

\begin{table*}
\centering{
         
\caption{\label{t5percentper2} List of lines with $\tau$ $>$ 0.1 in NGC 4636 using best-fit parameters from Table \ref{t:ngc}.}
  }
\begin{tabular}{c|c|c}
\hline
Lines &  Transition energy (keV) & Optical depth  \\

\hline
 O VIII & 0.6537 & 0.20 \\
Fe XVII& 0.8124 & 0.23\\
Fe XVII  & 0.8258 & 0.90  \\
Fe XVIII &0.8441& 0.18\\
Fe XVIII& 0.8512 & 0.41\\
Fe XVIII &0.8644&0.27\\
Fe XVIII & 0.8726 &0.24\\
Fe XVIII &0.8763&0.19\\
Fe XVIII &0.8772&0.17\\
Fe XVIII &0.8775& 0.22\\
Fe XVIII &0.8821&0.30\\
Fe XIX &0.9172&0.17\\
Fe XIX &0.9186&0.25\\
Fe XIX &  0.9229 & 0.31\\
Fe XVII & 1.0108& 0.12\\
Ne X & 1.0220 &0.13\\
Fe XVII & 1.0226&0.13\\
Mg XI & 1.3522&0.15\\
Si XIII & 1.8560& 0.24\\
\hline
\end{tabular}
 \end{table*}

\begin{table*}
\centering{
         
\caption{\label{t5percentper} List of lines with $\tau$ $>$ 0.1 scattering in  Perseus using best-fit parameters from Table \ref{t:xspec}.}
  }
\begin{tabular}{c|c|c}
\hline
Lines &  Energy (keV) & Optical depth  \\

\hline

Fe XXIV &  0.3817 & 2.93  \\
Fe XXIV &  0.3829 & 3.05 \\
Fe XXIV &  0.4032 & 0.78  \\
Fe XXIV &  1.1094  & 0.64 \\
Fe XXIV &  1.1242 & 0.70 \\
Fe XXIV &  1.1628 & 0.13   \\
Fe XXIV &  1.1676 & 0.24  \\
Fe XXV & 6.7004 & 1.00 \\
Fe XXV & 7.8815 & 0.16 \\

\hline
\end{tabular}
 \end{table*}

\begin{table*}
\centering{


\caption{\label{t:ngc}Best-fit parameters for for m$_{1}$(=\texttt{TBabs $\times$ bapec)}) and m$_{2}$(=\texttt{TBabs $\times$ rsapec}) fitting the summed
first-order RGS spectrum of NGC 4636 within the energy range 0.33-2.5 keV. The \texttt{XSPEC} convolution model \texttt{rgsxsrc} has been used to account for the spectral broadening of the extended source. The redshift is taken from \citet{2000MNRAS.313..469S} and has been frozen.}
  }
\begin{tabular}{c|c|c}
\hline
Parameter & m$_{1}$  & m$_{2}$  \\

\hline
nH(frozen, in units of $10^{22}$ atoms cm$^{-2}$) & 2e-2 & 2e-2  \\
kT (keV) & 0.64  $^{+0.005}_{-0.006}$  & 0.65  $^{+0.005}_{-0.005}$  \\
Abundance (solar) & 0.45 $^{+0.03}_{-0.03}$ & 0.45 $^{+0.04}_{-0.03}$  \\
Redshift (frozen) & 0.003129 & 0.003129 \\
Velocity (km/s) & 153.35  $^{+62.06}_{-45.69}$ & 165.01  $^{+57.29}_{-49.23}$ \\
nL (cm$^{-2}$) & - & 8.12$^{+0.08}_{-0.07}$ $\times$ 10$^{20}$ \\
norm &  2.02$^{+0.14}_{-0.11}$$\times$ 10$^{-3}$   & 2.08$^{+0.12}_{-0.13}$$\times$ 10$^{-3}$  \\
$\chi^{2}$ &  3847.43/3222 &  3798.39 /3221 \\
\hline
\end{tabular}
 \end{table*}

\begin{table*}
\centering{    
\caption{\label{t:xspec}Best-fit parameters for m$_{1}$(\texttt{TBabs $\times$ bapec})and m$_{2}$(\texttt{TBabs $\times$ rsapec}) fitting the  Fe K$\alpha$ complex of Perseus core observed by \textit{Hitomi}.
Fits are done for the energy range 6.0 to 7.0 keV for the outer region of Perseus core. The redshift is taken from \citet{2020A&A...633A..42S} and has been frozen.  }
  }
\begin{tabular}{l|r|r}
\hline
Parameter & m$_{1}$  & m$_{2}$  \\

\hline
nH(frozen, in units of $10^{22}$ atoms cm$^{-2}$) & 0.138 & 0.138  \\
kT (keV) & 4.07$\pm$ 0.02  & 4.08$\pm$ 0.02 \\
Abundance (solar) & 0.40 $\pm$0.001 & 0.39$\pm$0.002 \\
Redshift (frozen) & 1.7239E-02 & 1.72391E-02 \\
Velocity & 167.80$\pm$ 9.83 & 172.12$\pm$ 9.11 \\
 nL (cm$^{-2}$) & - &  (1.22$\pm$0.11) $\times$ 10$^{22}$   \\
norm & 0.93$\pm$ 0.01 & 0.96$\pm$0.02 \\
Cstat/d.o.f & 2390.51/1995 &  2314.05 /1994\\

\hline
\end{tabular}
 \end{table*}


\section{Model parameters:}\label{s:3}

The new resonant scattering model \texttt{rsapec} presented in this paper has been designed based on the modified version of \texttt{apec}, which models X-ray emission  from collisionally ionized, optically-thin  astrophysical plasmas \citep{2001ApJ...556L..91S, 2012ApJ...756..128F}. The \texttt{apec} model calculates the line emissivities using  three external model parameters-  electron temperature, metal abundances, and redshift.
However, solving for equations 1 and 2 is a  prerequisite for accurately synthesizing the spectrum for an optically thick plasma subject to resonance scattering. For constructing \texttt{rsapec}, we added two additional model parameters to \texttt{apec} - line of sight velocity broadening
in one dimension ($V_{1,turb}$) and $nL$, where  $nL$ is the electron density ($n_{e}$) integrated over  line-of-sight:
\begin{equation}\label{e:gamma}
 nL = \int_{{0}}^\infty n_{e} dr
\end{equation}

We chose $nL$ as a free parameter in \texttt{rsapec} as $n_{p}$ (and therefore $n_{e}$ \footnote{Assuming the majority of the free electrons in a collisionally-ionized cloud come from the ionization of hydrogen and helium, the following approximation was used: $n_{e}$ $\sim$ 1.2 $n_{p}$.}) is the only direct r-dependent quantity in equation \ref{e:gamma1}. The quantity $\delta_{i}$ depends on the temperature, which in galaxy clusters is only slightly dependent on radius \citep{2008MNRAS.390.1207R}, at least up to 100 kpc after which resonance suppression usually becomes unimportant. We, therefore, neglect the radial dependence of $\delta_{i}$ for simplicity.


We also introduced \texttt{rsvapec} and \texttt{rsvvapec}, equivalent to \texttt{vapec} and \texttt{vvapec}, respectively. \texttt{rsvapec} has 14 extra model parameters with the option to change abundances of the elements: H, He, C, N, O, Ne, Mg, Al, Si, S, Ar, Ca, Fe, Ni. \texttt{rsvvapec} has 30 extra model parameters with the option to vary the abundances of all the elements
between H and Zn.

We also adjusted the above models  to be compatible with pyXSPEC \citep{1996ASPC..101...17A}. 
Equations \ref{e:gamma1}- \ref{e:gamma2} are  simultaneously solved along with the $f$ factor and fitted with the observed spectrum within the AtomDB database to determine best-fit parameters for galaxies and clusters of galaxies
with optically thick emission lines.
Utilizing the fitted $nL$ value, abundances, and ionization fraction\footnote{ obtained from pyatomdb commands:
sess = pyatomdb.spectrum.CIESession(), ionfrac=sess.ionfraction(Temperature, Element, teunit=`K')}, it is also possible to get constraints on turbulent velocities.

\section{Model applications:}\label{s:4}

We apply our new RS model to model the resonance line suppression in NGC 4636 observed by RGS onboard \textit{XMM-Newton} and the outer core of the Perseus cluster observed by \textit{Hitomi}.  The observational log of the above observations is listed in Table \ref{t:o1}. In both cases, 
best-fit results were
obtained by minimizing C-statistic
\citep{1979ApJ...228..939C}.  The parameter errors were measured with 1$\sigma$ confidence.
The velocity listed in tables \ref{t:ngc} and \ref{t:xspec} is only the one-dimensional turbulent velocity, which dominates over thermal broadening in clusters and galaxies. The widths of the emission lines within AtomDB is calculated considering both thermal and turbulent broadening.

\subsection{NGC 4636 }

NGC 4636 is a giant  elliptical galaxy shining brightly in the X-rays. High-resolution XMM-Newton observations of NGC 4636
revealed a wealth of characteristic soft X-ray emission lines from various ion stages of nitrogen, oxygen, neon, magnesium, and iron \citep{2002ApJ...579..600X}. We use the 64 ksec first-order RGS spectrum of NGC 4636 as a test case of the RS model on elliptical galaxies. 
We used the \textit{XMM-Newton} tool  \texttt{rgsproc} for RGS data reduction for a cross-dispersion region of 3.5' width. The first-order spectra from RGS1 and RGS2 was combined using the tool \texttt{rgscombine}.
Table \ref{t:o1} lists the observational log of the spectra.
Several detected has $\tau$ $>$ 0.1, Fe XVII 15.015
\AA~(0.8258 keV) line being the only transition with significant optical depth ($\tau$ $\sim$ 0.93).  Refer to table \ref{t5percentper2} for the list of optical depths with more than $\tau$ $>$ 0.1.

Firstly, we attempted to fit the observed spectrum within the energy range 0.33-2.5 keV, including most of the line and continuum emission with a single-temperature \texttt{bapec} model multiplied with the pyXSPEC routine {\tt tbabs} to account for the galactic absorption. 
The absorbing hydrogen column density is set to 2.0$\times$10$^{20}$ cm$^{-2}$ using the  measurement of the galactic atomic hydrogen column density by \citet{2016A&A...594A.116H}.
The pyXSPEC routine {\tt rgsxsrc} was used to account for the spatial distribution of the X-ray emission along with the MOS1 image as done by \citet{2022MNRAS.514.4222F}. The boresight required in the {\tt rgsxsrc} routine was set  to the
X-ray brightest point of the image.
Using the model m$_{1}$(=\texttt{tbabs*bapec}) modified by \texttt{rgsxsrc} we obtained a best-fit temperature of $\sim$ 0.64 keV, and reduced $\chi$$^{2}$ = 3847.43/3222.
Next, we fit the observed spectrum within the same energy range with the \texttt{rsapec} model multiplied with {\tt tbabs} (m$_{2}$=\texttt{tbabs*rsapec}) and convoluted with {\tt rgsxsrc} . The fit to the spectrum was improved, with  $\chi$$^{2}$ = 3798.39/3221 and a best-fit temperature of $\sim$ 0.65 keV was obtained. The measured value of the one-dimensional turbulent velocity was 153.35  $^{+62.06}_{-45.69}$ km/s and 165.01  $^{+57.29}_{-49.23}$ km/s for m$_{1}$ and m$_{2}$ respectively, consistent with the turbulence measurement in NGC 4636 by \citet{2017MNRAS.472.1659O} (121  $^{+51}_{-35}$ km/s).
 Table \ref{t:ngc} lists the  best-fit parameters for m$_{1}$ and m$_{2}$ in 0.33-2.5 keV energy range. The measured value of $nL$  was 8.12 $\times$ 10$^{20}$ cm$^{-2}$, which agrees well with the estimate of $nL$ for NGC4636 as shown in the appendix. 

The left panel of Figure \ref{f:2} compares the fit of m$_{1}$ and m$_{2}$ within the same energy range. 
The emissivity of the 15.015\AA~Fe XVII line with optical depth close to unity is significantly affected, and the emissivities of the 
other lines listed in table \ref{t5percentper2} are partially affected by resonance scattering.
As a result, we get a slightly better fit with m$_{2}$ compared to m$_{1}$. The right panel of Figure \ref{f:2} compares the two models within a narrow energy range (0.80-0.85 keV) around the 15.015\AA~Fe XVII
line. Within the narrow range, the fit of m$_{2}$ ($\chi$$^{2}$ = 120.97/85  is noticeably better compared to that of m$_{1}$ ($\chi$$^{2}$ =  146.23/86) with an  F-
test probability $<<$ 0.01. We estimate the line suppression in Fe XVII due to RS to be $\sim$ 1.24.
Two-temperature/multi-temperature models showed no noticeable improvement in the fit to the spectrum.

\subsection{The Perseus cluster}

Possible suppression in the 6.7 keV line of the He-like iron in the X-ray spectrum of Perseus cluster has long been a topic of interest \citep{1987SvAL...13....3G, 1999AN....320..283A, 2004MNRAS.347...29C, 2004ApJ...600..670G, 2006MNRAS.370...63S, 2008MNRAS.390.1207R, 2013MNRAS.435.3111Z}. 
However, given that the Fe K$\alpha$ spectral lines were unresolved, no compelling evidence of RS was detected in the observed spectra. For the first time, the Soft X-ray Spectrometer (SXS) onboard \textit{Hitomi} separated the resonance line in the Fe K$\alpha$ complex from other line components, detecting a suppression in the Fe XXV resonance (w) by a factor of $\sim$ 1.3 at the core of Perseus cluster.
An ad hoc technique using a negative Gaussian to compensate for the RS suppresion was used to estimate this factor alongside an optically thin APEC model. 

Our newly formulated models use theoretical scattering probability calculations to estimate the RS factor for the emergence of the Fe XXV resonance (w) line from a collisionally-ionized plasma as described in  section 2. Optical depths were generated using the APED database (version 3.0.9).
Table \ref{t5percentper} shows the list of transitions with $\tau$ $>$ 0.1 for the best-fit parameters at Perseus core. The model details for estimating best-fit parameters are discussed below. 

We extracted the spectra for the outer region of Perseus core (30-60 kpc) using HEAsoft\footnote{https://heasarc.gsfc.nasa.gov/docs/software/heasoft/} for observations with observation IDs listed in table \ref{t:o1}. The observations' event files were merged and filtered using Xselect. Four distinct NXB spectra were derived from the four event files using \texttt{sxsnxbgen} and then averaged via \texttt{mathpha}. RMF, exposure map, and ARF are produced using \texttt{sxsmkrmf}, \texttt{ahexpmap}, and \texttt{aharfgen}, respectively. 

We fitted the extracted  without and with RS models: \texttt{bapec} and  \texttt{rsapec} around the Fe K$\alpha$ complex. The multiplicative pyXSPEC routine {\tt tbabs} was included in each fit to account for the galactic absorption, with the absorbing hydrogen column density set to 1.38$\times$10$^{21}$ cm$^{-2}$ \citep{2005A&A...440..775K} towards the direction of the Perseus cluster.

Table \ref{t:xspec}
compares the best-fit model  parameters fitting the observed spectrum  between m$_{1}$(=\texttt{tbabs*bapec}), and m$_{2}$(=\texttt{tbabs*rsapec}) within the energy range 6.0-7.0 keV.
Using m$_{1}$, we obtained a best-fit temperature of 4.07 keV with a \textit{Cstat/d.o.f} of 2390.51/1995.
The model m$_{2}$ improved the fit to the observed spectrum with a best-fit temperature of 4.08 keV and 
a \textit{Cstat/d.o.f} of 2314.05 /1994. The improvement to the fit is prompted by inclusion of the RS effect in the w line in m$_{2}$, the only optically thick line observed in the spectrum. The emissivity for the w line is noticeably suppressed by RS. The measured value of nL  was 1.2 $\times$ 10$^{22}$ cm$^{-2}$, which agrees well with the estimate of $nL$ for the Perseus cluster as shown in the appendix. 
The turbulent velocity was measured at 167.80 $\pm$ 9.83 km/s for m$_{1}$ and 172.12 $\pm$ 9.11 km/s for m$_{2}$, consistent with the value previously reported by \citet{2018PASJ...70...10H} (164 $\pm$ 10 km/s).


The left panel of Figure \ref{f:2} shows the spectrum of the Perseus outer core within 6.0-7.0 keV fitted with m$_{1}$ (\textit{Cstat/d.o.f} = 2390.51 /1995) and m$_{2}$ (\textit{Cstat/d.o.f} = 2314.05/1994). 
The right panel compares the two models within a narrower energy range 
immediately around  the Fe K$\alpha$ complex (6.5-6.6 keV). The fit of m$_{2}$ (\textit{Cstat/d.o.f} = 225.57/195) is significantly better than that of m$_{1}$ (\textit{Cstat/d.o.f} = 280.18/196) with the F-test probability of $<<$ 0.01, especially the fit to the w line. Comparing m$_{1}$ with m$_{2}$ we find a  RS line suppression factor of  $\sim$ 1.30 in w.

\section{Discussion}

We note that \citet{2006MNRAS.370...63S} constructed an \texttt{xspec} model for fitting spectra from clusters including resonance scattering. Their model calculated the emission spectrum from N annuli  divided into multiple subannuli in the sky using the APED atomic database (version 1.3.1) to generate
a list of resonance lines.

Our RS model \texttt{rsapec} is self-consistently built into the  AtomDB databse where the emission spectrum is estimated 
solving thousands of atomic rates, just like \texttt{apec/bapec} but inclusive of the RS effects. \texttt{rsapec} is almost as fast as \texttt{apec}, and can be used to model resonance scattering in both elliptical galaxies and clusters of galaxies. 
In the current
version of \texttt{rsapec}, scattering effects
are estimeted 
based on a spherical geometry as described in \citet{1998ApJ...497..587S}, which used a gas density profile with a 1D-$\beta$ model
with $\beta$ = 0.5. In clusters, \texttt{rsapec} will model resonance scattering for the inner 100 kpc radius, beyond which resonance enhancement due to scattering into the line of sight becomes important. In future, we plan to extend the model to include a full radiative transfer
treatment of the resonance scattering, which will include the flexibility to adjust the $\beta$ 
values,  resonance enhancement  at radial distances larger than 100 kpc, and non-spherical geometries. The current model will also be used as a template for modeling RS processes in sources like cataclysmic variables  for which the geometry or spatial resolution might differ but the approach to RS is the same.


We applied \texttt{rsapec} to fit the spectrum of the elliptical galaxy NGC 4636 and the Perseus clusters of galaxies. In perseus cluster, we estimated a suppression of $\sim$ 1.30 in the optically thick line Fe XXV K$\alpha$ w  (6.7004 keV) line, which compares well with the previously measured RS factor ($\sim$ 1.28) by \citet{2018PASJ...70...10H}. For NGC4636, we estimated a suppression factor of $\sim$ 1.24 in the Fe XVII (0.8258 keV) line, without any prior studies available.
We ignored all lines with $\tau$ $<$ 0.05  from our resonance scattering treatment.

The \textit{XRISM} mission, scheduled to launch in the next few months, will observe a sample of galaxies and clusters with optically thick lines. The \texttt{rsapec} model will be a self-consistent and  one-step tool  for extracting the best-fit parameters from the observed \textit{XRISM} spectrum fitting the optically thin and optically thick lines simultaneously in place of the previously applied ad-hoc technique of using negative Gaussians alongside optically thin \texttt{apec} model to account for the RS suppression.




\appendix

\section{Estimate for \lowercase{$n}L$ -- NGC 4636}

\citet{2015ApJ...803L..21V} estimated the radial profiles of electron density ($n_{e}$) of a sample of elliptical galaxies, including NGC 4636. Within the central 1 kpc region, $n_{e}$  is approximately constant in NGC 4636 ($n_{e}$ $\sim$ 0.12 cm$^{-3}$, refer to figure 1 in their paper), whereas within 
1-10 kpc, the density profile steeply declines. The density profile integrated over the inner 10 kpc is:

\begin{equation}\label{eqn:nL1}
\begin{split}
 \int_{{0}}^{10  \rm kpc} n_{e} dr & = \int_{{0}}^{1 \rm kpc} n_{e} dr +\int_{{1 \rm kpc}}^{10  \rm kpc} n_{e} dr\\ 
 & = 3.70 \times 10^{20}   \rm cm^{-2} + 4.62 \times 10^{20}    \rm cm^{-2} \\ &= 8.32 \times 10^{20}      \rm cm^{-2}
\end{split}
\end{equation}
 
 \section{Estimate for \lowercase{$n}L$  -- the perseus cluster}

We can get an estimate of $nL$ for the Perseus cluster using the radial electron density profile reported by \citet{2018PASJ...70...10H} (up to 1000 kpc):
\begin{equation}\label{eqn:nL1}
\begin{split}
 \int_{{0}}^{1000  \rm kpc} n_{e} dr & = 1.11 \times 10^{22}      \rm cm^{-2}
\end{split}
\end{equation}

The \texttt{rsapec} model is now available at the main AtomDB website: http://www.atomdb.org/.

\section*{Acknowledgment}
We acknowledge support by NASA ADAP Grant 80NSSC21K0637.

\bibliography{sample631}{}
\bibliographystyle{aasjournal}



\end{document}